\documentclass[onecolumn]{emulateapj}
\newcommand{\HST}{\emph{HST}}

\newcommand{\spitzer}{\emph{Spitzer}}
\newcommand{\tf}{24$\mu$m}
\newcommand{\mj}{$\mu$Jy}
\def \ellipse {{\sc Ellipse\ }}
\def \stsdas {{\sc Stsdas\ }}
\slugcomment{version: 9/1/05}

\begin{document}

\title{Optical Morphology Evolution of Infrared Luminous Galaxies in GOODS-N}
\author{J. Melbourne \altaffilmark{1}, D. C. Koo \altaffilmark{1}, E. Le Floc'h \altaffilmark{2,3}}

\altaffiltext{1} {University of California Observatories/Lick Observatory, Department of Astronomy and Astrophysics, University of California at Santa Cruz, 1156 High Street,  Santa Cruz, CA 95064.}
\altaffiltext{2}{Steward Observatory, University of Arizona, Tuscon, AZ 85721.}
\altaffiltext{3}{Associated to Observatoire de Paris, GEPI, 92195 Meudon, France}

\begin{abstract}
We combine optical morphologies and photometry from \HST, redshifts from Keck, and mid-infrared luminosities from \spitzer\ for an optically selected sample of $\sim800$ galaxies in GOODS-N to track morphology evolution of infrared luminous galaxies (LIRGs) since redshift $z=1$.  We find a 50\% decline in the number of LIRGs from $z\sim1$ to lower redshift, in agreement with previous studies.  In addition, there is evidence for a morphological evolution of the populations of LIRGs.  Above $z=0.5$, roughly half of all LIRGs are spiral, the peculiar/irregular to spiral ratio is $\sim0.7$, and both classes span a similar range of  $L_{IR}$ and $M_B$.  At low-$z$, spirals account for one-third of LIRGs, the peculiar to spiral fraction rises to 1.3, and for a given $M_B$ spirals tend to have lower IR luminosity than peculiars. Only a few percent of LIRGs at any redshift are red early-type galaxies.  For blue galaxies ($U-B < 0.2$), $M_B$ is well correlated with log($L_{IR}$) with an RMS scatter (about a bivariate linear fit) of $\sim0.25$ dex in IR luminosity.   Among blue galaxies that are brighter than $M_B = -21$, 75\% are LIRGs, regardless of redshift. These results can be explained by a scenario in which at high-$z$, most large spirals experience an elevated star formation rate as LIRGs.  Gas consumption results in a decline of LIRGs, especially in spirals, to lower redshifts. 

\end{abstract}


\section{Introduction}
Aussel et al. (1999) demonstrated that the large density of mid-IR(MIR) sources detected in deep ISOCAM (Cesarsky et al. 1996) data are dominated by galaxies at intermediate redshifts.  Luminous infrared galaxies (LIRGs), though rare locally, are common in the past and suggest a previously overlooked site of significant star formation, dust enshrouded starbursts.   Zheng et al. (2004), analyzed HST/WFPC2 images of 36 distant ($0.4 < z < 1.2$) ISOCAM detected LIRGs.  They classified 36\% of their objects as disk galaxies, 22\% as irregulars, 17\% as obvious mergers and 25\% compact sources.  They found a size-color relationship for the cores of their galaxies. Compact sources tend towards blue cores while the larger disk galaxies tend to have redder cores.  This suggested to the authors an evolutionary sequence of disk galaxy formation, with compact blue bulges forming first, followed by inside-out disk growth while the core reddens.  This theme was elaborated by Hammer et al. (2005; hereafter H05), who suggested that disk mergers are driving the evolution of both LIRGs and star formation over the last 8 Gyr.   In their picture, disk merger and rebuilding account for the various morphological stages seen in LIRGs at intermediate redshifts. For instance, a brief peculiar phase of heightened star formation is followed by a compact phase with dust and gas funneled into the center, growing the bulge.  Subsequent settling of gas into the potential well re-grow the disk from the inside out.   The authors suggest that this scenario may have occurred in 75\% of intermediate mass spirals since $z =1$.

ISOCAM data are only sensitive to the brightest MIR  sources at intermediate redshifts.  With the launch of the \spitzer\ space telescope (Werner et al. 2004), much deeper IR data have become available.  Using MIPS (Rieke et al. 2004) \tf\ observations from \spitzer, Bell et al. (2005; hereafter BELL05 ) study 1500 $z\sim0.7$ galaxies with photometric redshifts in the Chandra Deep Field South.  They found a somewhat different morphological make-up of LIRGs than the Zheng et al. and H05 studies and drew different conclusions.   BELL05 found more than half of LIRGS are undisturbed, massive ($M_{star} > 2 \times 10^{10} M_{\sun}$) spirals and less than a third strongly interacting.  They conclude that major mergers are not the primary factor in LIRG production or of the rapid decline of star formation since $z\sim0.7$.  They suggest instead a combination of gas depletion and minor mergers.

In this letter, we focus on one of the deepest MIR survey fields yet taken, the MIPS \tf\ image (Dickinson in preparation) of the Great Observatories Origins Deep Survey (GOODS, Giavalisco et al. 2004) north field.  Very deep optical \emph{Hubble Space Telescope} \ (\HST) imaging ($\sim 0.5$  mag deeper than BELL05) are used as well as spectroscopic redshifts from the Team Keck Redshift Survey (TKRS, Wirth et al. 2004).  We compare optical morphologies and luminosities  with IR luminosities of $\sim800$ optically selected galaxies to trace the morphological evolution of star forming galaxies since $z=1$.   The optically selected sample is  less biased against low IR luminosity objects than the previous studies. In addition, the deeper \HST\ data allow for more accurate morphological typing, especially of  fainter tidal features from mergers.  Our sample contains 3 times the number of LIRGs as the Zheng et al (2004) and H05 sample, and spans a wide range in redshift encompassing the BELL05 narrow redshift slice.    Section 3 explores the relationship between optical and IR luminosity, and its dependence on optical color, morphology and redshift.  Specifically we 1) look for evolution in the number density and morphological make-up of LIRGs since $z=1$; and  2) investigate the utility of optical color and luminosity as a predictor of IR luminosity. 
Section 4 synthesizes the results into a picture of LIRG evolution over the last 8 Gyr, similar to that of BELL05.

 We adopt Vega magnitudes and a flat, $h=0.7$, $\Omega_m=0.3$ cosmology throughout.
 
      
\section{The Data}
\subsection{Optical}
Galaxies are identified in the GOODS-N field (Giavalisco et al. 2004) using the $B,V,i$, and $z$ \HST\ images.  Spectroscopic redshifts are from TKRS (Wirth et al 2004), a magnitude limited ($R_{AB} \le 24.4$) survey of  2018 objects in GOODS-N.   TKRS is 53\% complete at its limiting magnitude (Wirth et al. 2004).  Elliptical aperture photometry from the \stsdas program \ellipse was generated from the 4 \HST\ bands for all objects in TKRS (J. Melbourne, in preparation).  Total magnitudes were calculated from the \ellipse measurements using a curve of growth technique. Following the prescription in Wilmer et al. (2005), k-corrections convert apparent magnitudes to absolute $M_B$ and rest-frame color $(U-B)$.  Our sample contains the 789 TKRS galaxies brighter than $M_B < -19.0$ grouped into three equal volume bins between $0.1 < z < 1.0$. This introduces a slight bias for redshifts  $z > 0.9$, where the TKRS limiting rest-frame magnitude is greater than $M_B \sim -19.0$.  This should not have a major impact on the LIRG detection as the majority of LIRGs are brighter than this limit (see for instance Figure 10 in Le Floc'h et al. 2005).

Galaxies are typed into one of 4 morphological categories, spiral, peculiar/irregular, compact and early-types (E/S0). Spirals, peculiars and early-types are classified by eye. Spirals exhibit symmetry, signs of bulge and disk components and spiral arms.  Peculiars are characterized by asymmetry, tidal features or obvious mergers.  Ellipticals are picked out as smooth, symmetric and red and lacking an obvious disk component.  Compacts are selected by having half-light radii smaller than 3 kpc.  At low and intermediate redshifts ($z < 0.8$) the typing is done with the very deep $V$- and $i$-band images.  At higher redshift, the $z$-band image (rest-frame $M_B$) is used to account for the morphological k-correction.   The typing was done by three independent researchers with 70\% full agreement, and 100\% agreement between at least two of three classifications.  A final classification was assigned by majority.  As a check of our typing, we classify the 27 non-compact objects in Zeng et al. (2004) and agree with all but one of the classifications.  Furthermore we clone our low redshift LIRGs to the pixel scale and signal to noise of the z=0.95 LIRGs and find that the classifications are not affected.    


\begin{figure}
\includegraphics{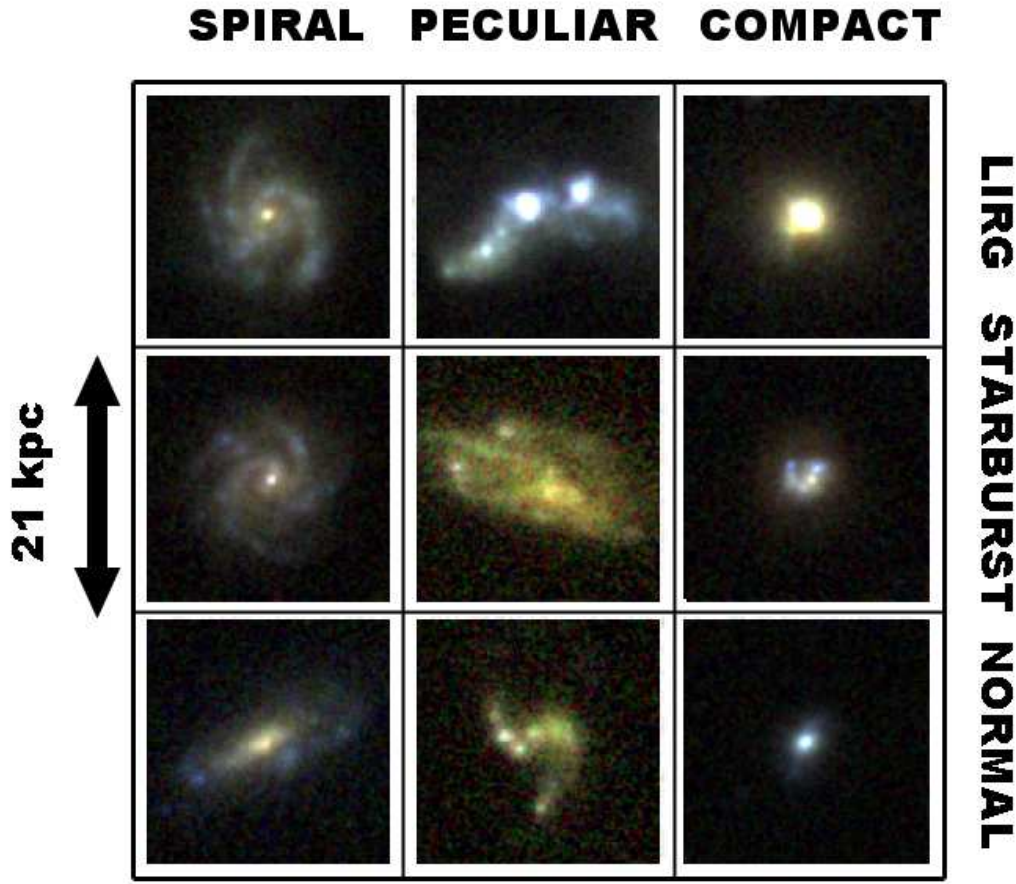}[h]
\centering
\figcaption{\label{fig:images} ACS 3 color images ($Viz$-band)  of 9 blue galaxies at $z\sim0.7$ are shown.  The images grouped into 3 categories; spirals, peculiars/ irregulars, and compacts.  The first two categories are grouped by visual inspection, while compacts are objects with half light radii smaller than 3 kpc.  Three objects from each morphological class are shown; one LIRG, one starburst, and one 'normal' galaxy.  Morphology alone is not enough to predict the IR luminosity of blue objects, though early-type red galaxies are very rarely LIRGs.}
\end{figure}

\subsection{\tf\ Fluxes}
\tf\ fluxes are measured from the publicly released MIPS mosaic image of the GOODS-N field (GOODS DR1+ data release, M. Dickinson, in preparation).  The image is flux calibrated and aligned to the astrometry of the IRAC images of the same region (Chary et al. 2004).  The MIPS \tf\ point-spread-function (PSF) has a full-width-half-max (FWHM) of $\sim6\arcsec$, meaning virtually every source on the image is point-like.  As a result, our photometry techniques are borrowed from those used in crowded star fields.  First we generate a PSF for the image by median combining 10 flux scaled isolated sources.  Then for each TKRS galaxy position, we 1) subtract all neighbors with centroids separated by more than $2\arcsec$ and less than $75\arcsec$ from the location of interest. 2)  Measure the flux in a $9.8 \arcsec$ aperture at the location of interest.  3) Apply an aperture correction of 1.34 to the flux measurement (taken from the MIPS Data handbook V2.1), to account for the large amount of flux in the wings of MIPS PSF.  4) Apply an additional correction to the flux such that $image - (scaled\; PSF)$ is within 0.5 \% of the sky value.

For bright objects ( $> 200$ \mj), this technique reproduces the results in the DR1+ bright source catalogue (R. Chary, in preparation) to within 5\%.  For fainter sources ($100 < f < 200$ \mj), the results vary with an RMS of 15\%. While the DR1+ catalogue supplies measurements of objects brighter than 100 \mj, the measurement strategy outlined above provides flux estimates of faint sources to 25 \mj.  In addition, because we select galaxies in the optical, we measure \tf\ fluxes for objects that may not have been identified by the standard DAOphot source finding routines.  

Elbaz et al. (2002) demonstrate that the MIR is well correlated with total IR luminosity, even at cosmological distances.  We estimate the total IR luminosity, $L_{IR}(8-1000 \mu m),$ from the MIPS \tf\ flux following the prescription in  Le Floc'h et al. (2005).  For each redshift and \tf\ flux, this method derives the monochromatic luminosity at the corresponding rest-frame wavelength, which is then translated into a total IR luminosity ($8-1000 \mu$m) according to libraries of IR spectral energy distributions (SEDs). The SEDs considered in this work were taken from Chary \& Elbaz (2001).  Other templates give consistent results within 0.3 dex,  which is also considered to be the accuracy of our estimates.  Following the convention established in Sanders \& Mirabel (1996), LIRGs are defined as log($L_{IR} / L_{\sun}) > 11$. Using the conversion to star formation rates (SFR) given in Kennicutt (1998), LIRGs have SFR $>  17M_{\sun}yr^{-1}$.  Starbursts are defined as $ 10 < log(L_{IR} / L_{\sun}) < 11$ and have SFR of 2-17 $M_{\sun}yr^{-1}$, while normal blue galaxies, log($L_{IR}/L_{\sun}) < 10$, have SFR less than 2 solar masses per year.  
 
\section{Results}

Figure \ref{fig:images} shows ACS 3-color images of sample galaxies from the intermediate redshift bin ($z\sim0.7$).  Galaxies are placed into 4 morphological categories, spiral, peculiar/irregular, compact and early-types (E/S0 not shown on image).   For each morphological category, the figure shows a LIRG, a starburst galaxy and a normal galaxy.  Aside from the dearth of red early-type LIRGs (only 5 of the 119 LIRGs are classified as ellipticals), morphology is not a good predictor of IR luminosity.

\begin{figure}
\centering
\includegraphics[scale=0.85]{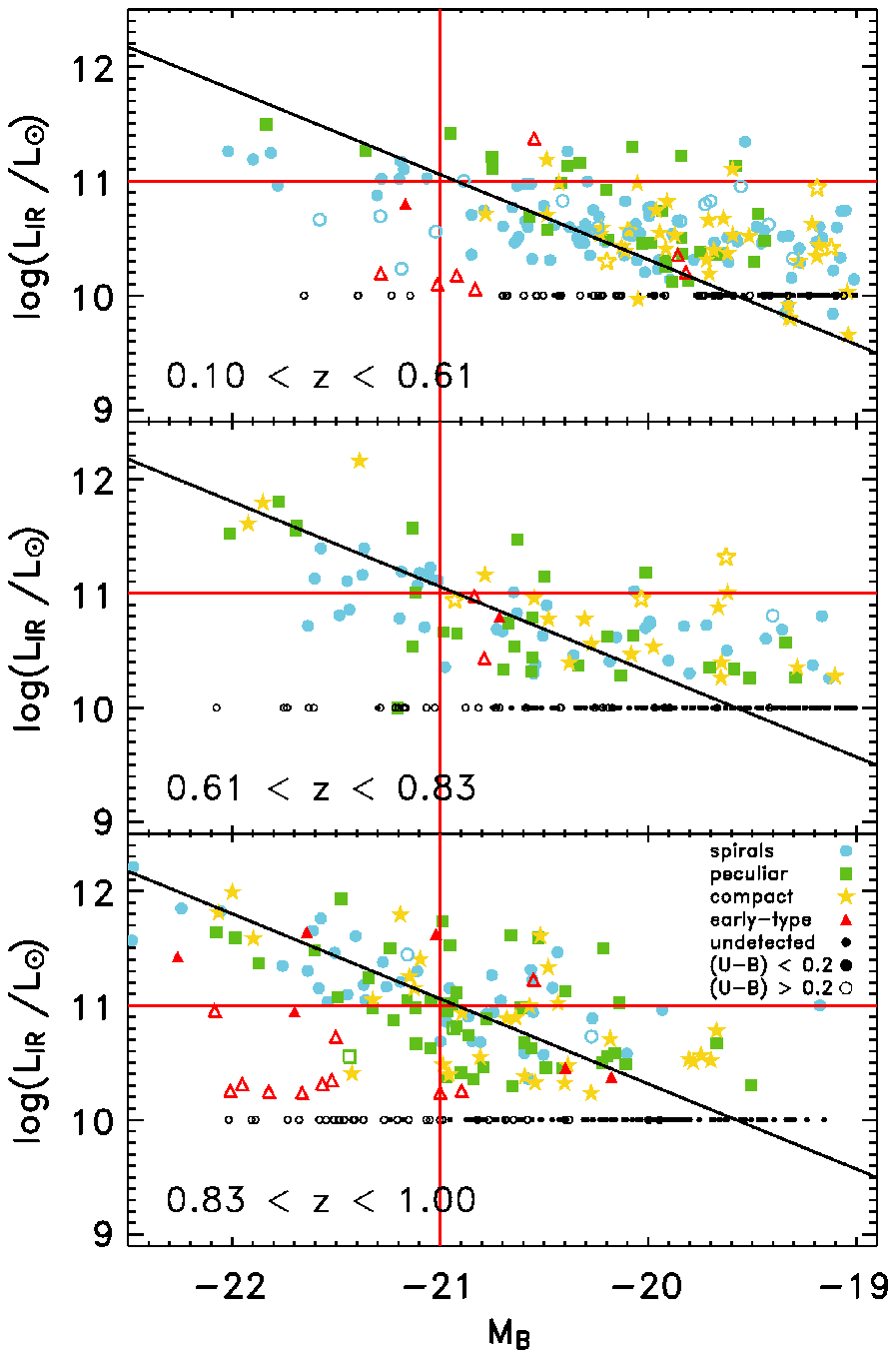}
\figcaption{\label{fig:MBIR} A plot of  total IR luminosity (in solar units) vs. rest-frame $B$-band magnitude in three equal volume redshift bins.  LIRGs lie above the horizontal red line.  Morphology is indicated by color and shape, spirals are blue circles, peculiar/irregular are green squares, compacts are yellow stars and early-types (E/S0) are red triangles.  Objects undetected in the MIPS image are shown as small black circles.  Filled symbols are blue  galaxies $(U-B) < 0.2$ and open symbols are galaxies redder than this limit. For blue galaxies there is a clear trend of increasing IR luminosity with $M_B$.  The RMS scatter in IR luminosity about the bivariate least squares fit (black line) to the entire sample of blue galaxies is $\sim 0.25$ dex.   Brighter than $M_B <-21$ (left of the red vertical line), 75\% of galaxies are LIRGs, regardless of redshift.  In the highest redshift bin spirals and peculiars are well mixed.  In contrast, at a given $M_B$, in the low-$z$ bin, spirals tend to have lower IR luminosities than peculiars. }
\end{figure}


Table \ref{table:LIRGmorph} reports the morphological typing of the 119 LIRGs binned into 3 equal volume bins (low, intermediate and high redshift) and 1 additional low redshift bin to highlight morphology evolution of the sample. Two major conclusions can be drawn from these numbers.  First, the number density of LIRGs has dropped by 50\% from the high redshift bin to the intermediate redshift bin, and continues to drop in the low redshift bins.  Second the morphological fraction of the LIRGs has changed.  Above $z=0.5$ the peculiar to spiral ratio for LIRGS is $p/s\sim0.7$, whereas below $z=0.5$ the peculiar to spiral ratio is $p/s = 9/7=1.3$.  A quick Monte Carlo simulation indicates that drawing such a skewed sample from a distribution with $p/s=0.7$ is unlikely at the 84\% level, for this small sample size.  We caution that the reclassification of 4 low z objects would be enough to change these fractions to match the higher $z$ results.  Therefore, larger samples of low-$z$ LIRGs will be helpful in verifying if this trend is real.  While the result is marginal, Figure \ref{fig:MBIR}, a plot of IR luminosity vs. $M_B$ for the entire sample of galaxies, provides additional evidence for a change in LIRG population with redshift.

\begin{deluxetable}{rcc|cc|cc|cc}
\tablewidth{0pt}
\tablecaption{Morphology of TKRS LIRGs \label{table:LIRGmorph}}
\tablehead{
\colhead{} &
\multicolumn{2}{l|}{$0.83 < z < 1.00$} &
\multicolumn{2}{l|}{$0.61 < z < 0.83$} &
\multicolumn{2}{l|}{$0.10 < z < 0.61$} &
\multicolumn{2}{l}{$0.10 < z < 0.50$} 
}  
\startdata
&   \# &  \%&   \#  & \%&   \# &  \%&   \#  & \% \\
total &		63 & 100 &30 & 100 &25 & 100 & 18	&100 \\  
spiral &	29 & 46 & 14 & 47  &12 & 48  & 7	&38    \\
peculiar&	20 & 32 & 10 & 33  &10 & 40  & 9	&50    \\
compact	&       10 & 16 & 6  &20  & 2  &8  & 2&	11    \\
elliptical &	4  & 6 &  0  & 0   & 1 & 4    & 0 &	0  \\
\tableline
\enddata   
\end{deluxetable}

Figure \ref{fig:MBIR} plots IR luminosity vs. $M_B$, in low, intermediate and high redshift equal volume bins. The red horizontal line divides starburst galaxies (below the line) and LIRG luminosity (above the line).  Objects with MIPS \tf\ measurements below $25 \mu$J are shown as black circles and are designated 'normal' galaxies  The morphologies of the starburst and LIRG samples are indicated: spiral galaxies are blue circles, peculiar/irregular are green squares, compact galaxies are yellow stars and ellipticals are red triangles.  Filled symbols are blue  galaxies $(U-B) < 0.2$ and open symbols are galaxies redder than this limit.  This plot illustrates the morphological evolution hinted at in the previous analysis.  In the high redshift bin, spirals and peculiars are well mixed, spanning similar $M_B$ and $L_{IR}$ ranges.  In the low redshift bin, the two populations segregate with the peculiars systematically at higher IR luminosity than the spirals.  This results in a change of the LIRG peculiar to spiral ratio.  

The morphological segregation of low-$z$ IR luminous galaxies has been observed by Ishida (2004) in a statistically complete sample of 56 local LIRGs drawn from IRAS Bright Galaxy Sample (Soifer et al. 1997).  The Ishida  study finds that 100\% of galaxies with log($L_{IR}) >11.5 $ show at least some evidence for tidal features.  The majority of galaxies with $11.1 <$ log($L_{IR}) <11.5$, also tend to be disturbed but $\sim30\%$ are bright isolated spirals.  The Ishida study indicates that at lower IR luminosities, spirals comprise a larger fraction of the total.    The implications of these morphological changes with redshift will be explored in the Discussion section.

  
While morphology does not appear to be a good indicator of IR luminosity for blue galaxies, $(U-B) < 0.2$, absolute $M_B$ magnitude is.  Blue galaxies (filled symbols in Figure \ref{fig:MBIR}) show a clear trend of increasing IR luminosity with $M_B$, regardless of morphology.   This trend, shown previously without a color-cut (Chary \& Elbaz 2001; Le Floc'h et al. 2005), was found to have a large scatter.  With a color-cut, this scatter is significantly reduced. For our sample, the RMS scatter in IR luminosity about the best-fit line (black line, a bivariate least-squares fit to the full sample of blue galaxies, used for illustrative purposes) drops from $\sim0.40 $ dex without the color-cut to $\sim0.25$ dex with the cut. This scatter is actually somewhat smaller than our roughly estimated IR luminosity error ($0.3$ dex).   Figure \ref{fig:MBIR} shows that both the optical and IR luminosities of the blue galaxies are increasing with redshift.  Roughly 75\% of blue galaxies brighter than the characteristic luminosity of  $M_B < -21.0$ (red vertical line) are LIRGs, regardless of redshift.  Only 3 of the 93 blue galaxies brighter than this limit are undetected in the MIPS image.  


\section{Discussion}
Studies of  galaxies in the UV and optical have shown that the average comoving star formation rate (SFR) density of the universe has declined by a factor of $\sim10$ since $z=1$ (Lilly et al. 1996; Madau et al. 1998; Hogg et al. 1998; among others), and observations in the sub mm wavelengths indicate that it may have peaked around $z=2$ (Chapman et al. 2005).  Because IR luminosity strongly correlates with SFR (Kennicutt 1998),  we see a decline, by $\sim50\%$ in the numbers of LIRGs from the high redshift bin ($z\sim0.95$)  to the  intermediate ($z\sim0.7$) and low ($z\sim0.4$)  redshift bins.  This drop in LIRG numbers matches well the results from Le Floc'h et al. (2005), who found that the comoving IR energy density evolves as $(1+z)^{3.9}$.  Assuming the drop in IR energy density were dominated by a reduction in the number of LIRGs, we expect a 50\% drop in the number of LIRGs of from $z=1$ to $z=0.7$, which is what we find.  Despite this decline, $\sim75\%$ of the brightest ($M_B < -21.0$),  blue ($(U-B) <  0.2$) galaxies are LIRGs, regardless of redshift.  Therefore it appears that the $B$-band optical and IR luminosities of galaxies are linked, with SFR as an obvious candidate for the connection.   

A large gas supply is required to sustain elevated SFR at the LIRG level.  A LIRG event in the Milky Way lasting only 100 Myr would use up the entire Galactic HI-disk gas supply (assuming $\sim 2.5 \times 10^{9} M_{\sun}$ of HI gas, Nakanishi \& Sofue 2003).  However, roughly half of the intermediate and high redshift LIRGs in our sample are comprised of large normal looking spirals like the Milky Way, the difference being that high-$z$ LIRG spirals can be as luminous in the IR as peculiar galaxies.  BELL05 finds that LIRG spirals tend to be massive with stellar masses $M_{\sun} > 2 \times 10^{10} M_{\sun}$. Figure 4 of BELL05 shows that the specific SFR of a handful of the most massive LIRGs are at or below their past average, indicating that either these galaxies experienced multiple LIRG events, or have had enough gas to continuously form stars at a LIRG level. Eventually this gas, if not resupplied, will be used up.   At low redshift, the make-up of the LIRG population changes.  LIRG spirals exhibit lower IR luminosities than their peculiar counterparts, and also become more rare.  Possible explanations for this include 1) the gas reservoirs in spirals are depleted with time, or 2) gas accretion is reduced with time.   With a depleted gas supply, and/or reduced gas accretion,  large local spirals can still form stars but generally at rates and below the LIRG level.  One way to build up a large, high-density gas supply at low-$z$ is to merge spirals together, an event revealed by peculiar morphology.  

This picture of gas depletion differs from that of H05, who suggest that major mergers of spirals are driving the star formation and creation of LIRGs since $z\sim1$.  In the H05 picture, most intermediate-mass spirals undergo at least one major merger since $z\sim1$, during which the morphology changes from spiral to peculiar (merger) to compact (bulge formation) to spiral again (gas infall reforms the disk).  If this scenario is responsible for most LIRGs, the process should maintain a roughly constant peculiar to spiral ratio,  as each morphological phase leads, proportionally in time, into the next.   Such a prediction is hard to reconcile  with both the large fraction of undisturbed spirals at high redshift and the evolutionary changes seen in the morphological fractions of LIRGs.

While the changing peculiar to spiral ratio may appear at face value to be incompatible with the H05 picture, we would like to caution the reader on several points.  First the evidence against the H05 picture is based on visual morphological classifications and thus somewhat subjective.  While identification of major mergers is relatively straight forward at low redshift, tidal features and other indications of peculiar morphology become harder to identify at higher redshift, as signal-to-noise drops. By using very deep optical imaging and cloning the low-$z$ objects to higher redshift, we have tried to minimize these concerns.  In the future, adoption of quantitative morphological indicators such as concentration,  asymmetry, clumpiness (Conselice 2003) and Gini/M$_{20}$ (Lotz, Primack \& Madau 2004)  may help alleviate these concerns.  Secondly,  larger samples preferably in other parts of the sky will help to account for cosmic variance and increase our relatively modest sample of low redshift LIRGs. Thirdly, gas depletion at low-$z$ may itself change the disk rebuilding efficiency in the H05 scenario and result in altered LIRG morphological fractions.  

\acknowledgments
We would like to thank Leonidas Moustakas, Mark Dickinson, and Ranga-Ram Chary, Catherine Ishida and Sandra Faber for helpful comments on the project.  This work has been supported in part by the NSF Science and Technology Center for Adaptive Optics managed by UC Santa Cruz under the cooperative agreement No. AST-9876783.

\end{document}